# Mutual information and swap operation in the two-qubit Heisenberg model with Dzyaloshinskii-Moriya anisotropic antisymmetric interaction

Guo-Feng Zhang*

*Department of Physics, School of Sciences, Beijing University of Aeronautics & Astronautics, Xueyuan Road No. 37, Beijing 100083, People's Republic of China*

**Abstract:** Mutual information and swap operation in the two-qubit Heisenberg model with the Dzyaloshinskii-Moriya (DM) anisotropic antisymmetric interaction are investigated. It is found that the mutual information *I* of such a quantum channel is sensitive to the initial conditions and declines with the increase of the temperature. The DM interaction can lead to a higher mutual information, especially for the ferromagnetic case. The entanglement quality of input states cannot enhance the mutual information of the quantum channel. When the DM interaction is large, the mutual information will be the maximum value 2 for a non-entangled input state, but it is 1 for a maximally entangled input one. It is also shown that the swap operation can be implemented for some kinds of DM coupling. The conditions of the DM coupling under which the swap operation is feasible are established.

## I. INTRODUCTION

The study of quantum information processing has opened up an exciting new field in which an intrinsic principle of quantum mechanics plays an important role [1]. For experimental investigations of quantum computation and quantum communication, the thermal equilibrium state in the Heisenberg model is one fundamental sort of mixed and entangled states. In some schemes of teleportation using the Heisenberg model, half of two qubits can be teleported by the sole two-qubit mixed state [2]. In other schemes, the entanglement of a special mixed state, i.e., the Werner state, can be transferred through a pair of two-qubit mixed states [3]. Also entanglement teleportation via thermal entangled states of a two-qubit Heisenberg *XX* chain has been reported [4]. Yeo *et al.* [5] studied the influence of anisotropy and magnetic field on quantum teleportation via a Heisenberg *XY* chain. We studied thermal entanglement and teleportation in a two-qubit Heisenberg chain with the Dzyaloshinskii-Moriya (DM) anisotropic antisymmetric interaction [6]. However, to see more clearly the possible applications of the Heisenberg model in quantum teleportation, the mutual information of the quantum channel also needs to be analyzed. Moreover, the minimal requirements for a quantum computer (QC) architecture are the existence of fundamental quantum bits and the ability to carry out qubit operations, such as the quantum exclusive-or gate [also known as controlled-not (CNOT)], the Walsh–Hadamard gate and the swap gate, which is defined by $U_{swap}|\Phi\rangle\otimes|\Psi\rangle=|\Psi\rangle\otimes|\Phi\rangle$ [7]. The swap operation is a particularly intriguing process; although it takes product states to product states, it is the most nonlocal operation and can act as a double-teleportation [8]. The swap gate itself is not universal; however, it can be proved that the square root of a swap gate $\sqrt{U_{swap}}$ is universal. A CNOT gate can be constructed through a combination of single-qubit operations and $\sqrt{U_{swap}}$ [9]. In the proposed spin-based QC architectures, the exchange interaction between spins plays a fundamental role in the establishing of two-qubit entanglement [10], while the Zeeman splitting, which is a function of the external magnetic field, provides various single qubit operations. The effects of anisotropic magnetic fields on the interplay between Zeeman splitting and the swap operation have been studied for the isotropic Heisenberg model [11, 12]. However, cases in which the anisotropic interaction exists have seldom been considered. Though Burkard *et al.* [13] and Bonesteel *et al.* [14] investigated the effect of spin-orbit effects on a quantum gate, the appropriate magnitude of spin-orbit coupling for implementing the swap gate has yet to be given.

* E-mail:gf1978zhang@buaa.edu.cn

In this paper, the mutual information and swap operation of the two-qubit Heisenberg model in the presence of the DM anisotropic antisymmetric interaction are investigated. The model will be given in Sec. II. In Sec. III. The mutual information will be investigated. The implement consideration of swap gate based on the model will be given in Sec. IV. In Sec. V a discussion concludes the paper.

## II. THE MODEL AND SOLUTIONS

Now we consider the Heisenberg model with the Dzyaloshinskii-Moriya (DM) interaction

$$H_{DM} = \frac{J}{2}[(\sigma_{1x}\sigma_{2x} + \sigma_{1y}\sigma_{2y} + \sigma_{1z}\sigma_{2z}) + \vec{D}\cdot(\vec{\sigma}_1 \times \vec{\sigma}_2)], \tag{1}$$

where $\vec{D}$ is the DM vector coupling. The DM anisotropic antisymmetric interaction arises from spin-orbit coupling [15, 16]. Where $J > 0$ corresponds to the antiferromagnetic case and $J < 0$ the ferromagnetic case, here we assume that $J$ is a rational number. To see the effect of the anisotropic parameter $\vec{D}$, we choose $\vec{D} = D\vec{z}$. Then the Hamiltonian $H_{DM}$ becomes

$$H_{DM} = J[(1+iD)\sigma_{1+}\sigma_{2-} + (1-iD)\sigma_{1-}\sigma_{2+}] + \frac{J}{2}\sigma_{1z}\sigma_{2z}. \tag{2}$$

The eigenvalues and eigenvectors of $H_{DM}$ are given by

$$H_{DM}\left|00\right\rangle = \frac{J}{2}\left|00\right\rangle, H_{DM}\left|11\right\rangle = \frac{J}{2}\left|11\right\rangle, H_{DM}\left|\pm\right\rangle = (\pm J\sqrt{1+D^2} - \frac{J}{2})\left|\pm\right\rangle, \tag{3}$$

where $\left|\pm\right\rangle = 1/\sqrt{2}(\left|01\right\rangle \pm e^{i\theta}\left|10\right\rangle)$ and $\theta = \arctan D$. As thermal fluctuation is introduced into the system, the state of a typical solid state system at thermal equilibrium (temperature *T*) is $\rho(T) = e^{-H/(kT)}/Z$, where *H* is the Hamiltonian and $Z = Tre^{-H/(kT)}$ is the partition function. In the standard basis {$\left|11\right\rangle, \left|10\right\rangle, \left|01\right\rangle, \left|00\right\rangle$}, the density matrix $\rho(T)$ can be expressed as

$$\rho(T) = \frac{1}{Z}\begin{pmatrix} e^{-\beta J/2} & 0 & 0 & 0 \\ 0 & e^{\beta(J-\delta)/2}(1+e^{\beta\delta})/2 & e^{i\theta}e^{\beta(J-\delta)/2}(1-e^{\beta\delta})/2 & 0 \\ 0 & e^{-i\theta}e^{\beta(J-\delta)/2}(1-e^{\beta\delta})/2 & e^{\beta(J-\delta)/2}(1+e^{\beta\delta})/2 & 0 \\ 0 & 0 & 0 & 0 \end{pmatrix}, \tag{4}$$

where $Z = 2e^{-\beta J/2}[1 + e^{\beta J}\cosh(\beta\delta/2)]$, $\beta = 1/(kT)$ and $\delta = 2J\sqrt{1+D^2}$. In the following calculation, we will write the Boltzmann constant *k*=1. Note that we are working in units where *D* and *J* are dimensionless, and the entanglement of two qubits can be measured by the concurrence *C* [17]. From Ref. [6], we know that When *D*=0, i.e.,$\delta = 2J$, the critical temperature above which the concurrence vanishes is $T_c = 2J/\ln 3$ for *J*>0, but there is no entanglement at any temperature for *J*<0.

Similar to standard teleportation, entanglement teleportation for the mixed channel of an input entangled state is destroyed and its replica state appears at the remote place after applying a local measurement in the form of linear operators. When a two-qubit state $\rho_{in}$ is teleported via the channel, the output state $\rho_{out}$ is [2]

$$\rho_{out} = \sum_{ij} p_{ij}(\sigma_i \otimes \sigma_i)\rho_{in}(\sigma_i \otimes \sigma_i), \tag{5}$$

with $\sigma_i (i = 0, x, y, z)$ signifying the unit matrix $I$ and three components of the Pauli matrix $\vec{\sigma}$, respectively, $p_{ij} = Tr[E^i \rho(T)]Tr[E^j \rho(T)]$, $\sum_{ij} p_{ij} = 1$ and $\rho_{in} = |\varphi_{in}\rangle\langle\varphi_{in}|$.

Here $E^0 = |\Psi^-\rangle\langle\Psi^-|, E^1 = |\Phi^-\rangle\langle\Phi^-|, E^2 = |\Phi^+\rangle\langle\Phi^+|, E^3 = |\Psi^+\rangle\langle\Psi^+|$,

in which $|\Psi^\pm\rangle = (1/\sqrt{2})(|01\rangle \pm |10\rangle)$, $|\Phi^\pm\rangle = (1/\sqrt{2})(|00\rangle \pm |11\rangle)$. Based on Ref. [17], we can get the concurrence of the output state $C_{out}$.

## III. MUTUAL INFORMATION OF THE QUANTUM CHANNEL

Compared with classical information theory, the mutual information $I$ in the quantum communication theory can imply the classical capacity of a quantum channel. For two-qubit input states, a mutual information value of 2.0 means that the classical information carried by the input states can be totally transmitted via the quantum channel. If the value of $I$ is 0, it means that the original classical information coded in the input states is totally destroyed after quantum teleportation. If the value of I is $0 < I < 2.0$, it means that the original classical information coded in input states is partially destroyed after quantum teleportation. Here we consider the following four input states:

$$\begin{aligned} |\varphi_{in}^1\rangle &= \cos\omega|00\rangle + \sin\omega|11\rangle, |\varphi_{in}^2\rangle = \sin\omega|00\rangle - \cos\omega|11\rangle, \\ |\varphi_{in}^3\rangle &= \cos\Gamma|01\rangle + \sin\Gamma|10\rangle, |\varphi_{in}^4\rangle = \sin\Gamma|01\rangle - \cos\Gamma|10\rangle, \end{aligned} \quad (6)$$

The parameters $\omega$ and $\Gamma$ describe two sets of entangled states with different amplitudes. We assume that the probability of the above state is the same, i.e., 1/4. Based on Refs. [18, 19], the mutual information can be written as

$$I = 2 - \frac{1}{4}\sum_l S(\rho_{out}^l), l = 1,2,3,4, \quad (7)$$

In Eq. (7), the Von Neumann entropy is $S(\rho) = -tr(\rho \log_2 \rho)$. According to Eqs. (5-7), we can get the mutual information of the quantum channel.

The mutual information $I$ is illustrated in Figs. 1-4. Figure 1 is a plot of the mutual information $I$ as a function of the input states ($\omega$, $\Gamma$). From the figures, we can see that the mutual information $I$ is sensitive to the initial conditions and evolves with respect to both $\omega$ and $\Gamma$. There is a minimum mutual information of $I$ =1.55 when the input signal states are maximally entangled with $\omega = \Gamma = (n + 1/4)\pi$ or $(n + 3/4)\pi$ (0,1,2,3,...), while when the input signal states are not entangled, the mutual information $I$ can reach a maximum value, which is 1.65. It is interesting to note that the entanglement quality of the input states cannot enhance the mutual information in entanglement teleportation. Comparing Fig. 1(a) with (b), we can see that the introduction of the DM interaction can make mutual information have the maximum value of 2 when the input states are not entangled; meanwhile, the DM interaction weakens the mutual information when the input states are maximally entangled (for example, $D$=0, $I$=1.55 for $\omega = \Gamma = \pi/4$, but $D$=5, $I$=1.00).

Figure 2 demonstrates the dependences of mutual information on $J$ and $T$ for different DM interactions when the input states are maximally entangled. It is shown that the mutual information $I$ of such a quantum channel declines (increases) with the increase of the temperature (spin-spin coupling $J$). The DM interaction can lead to higher mutual information for the ferromagnetic case, contrary to the antiferromagnetic case.

The mutual information $I$ is plotted in Fig. 3 for varying input states $\omega$ and $\Gamma$. It can be seen that the mutual information $I$ of maximally entangled states (solid line) decreases more rapidly than that of non-entangled input states (dotted line). From the left panel, we can see that there is almost no difference in $I$ between a maximally entangled input state and a non-entangled input state when the temperature is very small for $D$=0. However, the difference becomes pronounced as the DM interaction is introduced, which can be seen from the right panel.

We give a plot of I as a function of $D$ in Fig. 4. As the DM interaction increases, the mutual information $I$ decreases to 1 for a maximally entangled input state, but it increases to 2 for a non-entangled input state when $D$ is large. This is due to the fact that $C_{out} = 0$ when $D \to \infty$ for any input state. In other words, any one input state will be non-entangled when $D \to \infty$. Thus, the mutual information is 2 for a non-entangled input state, but it is 1 for a maximally entangled input state.

## IV. SWAP OPERATION IN THIS MODEL

In order to investigate the effect of the DM interaction on the swap operation, the initial state is chosen as a product state given by

$$\psi(0) = (\alpha_1 |1\rangle + \alpha_2 |0\rangle) \otimes (\beta_1 |1\rangle + \beta_2 |0\rangle) = \begin{pmatrix} \alpha_1 \\ \alpha_2 \end{pmatrix} \otimes \begin{pmatrix} \beta_1 \\ \beta_2 \end{pmatrix}, \tag{8}$$

and it then evolves under the Hamiltonian (1):

$$\psi(t) = e^{-iH_{DM}t} \psi(0). \tag{9}$$

If the wave function becomes $(\beta_1 |1\rangle + \beta_2 |0\rangle) \otimes (\alpha_1 |1\rangle + \alpha_2 |0\rangle)$ at some time, then the swap operation has been achieved. From the above expressions, a swap is achieved by exchanging the coefficients of the unpolarized state $|01\rangle$ and state $|10\rangle$. Expanding the initial state in the basis of eigenstates of Hamiltonian (1), Eq. (9) becomes

$$\psi(t) = \alpha_2 \beta_2 e^{-iJt/2} + \alpha_1 \beta_1 e^{-iJt/2} + \Delta_+ e^{-i(\delta - J)t/2} + \Delta_- e^{i(\delta + J)t/2}, \tag{10}$$

with $\Delta_\pm = \sqrt{2}(\alpha_2 \beta_1 \pm \alpha_1 \beta_2 e^{-i\theta})/2$. If a two-qubit system is in a disentangled state, the reduced density matrix of either spin is pure. The reduced density matrix of the fist spin is given by:

$$\begin{aligned} \rho_{1,11} &= |b_2|^2 + |b_4|^2, \rho_{1,00} = |b_1|^2 + |b_3|^2, \\ \rho_{1,10} &= b_1^* b_4 + b_2 b_3^*, \rho_{1,01} = b_1 b_4^* + b_2^* b_3, \end{aligned} \tag{11}$$

with $b_1 = \alpha_2 \beta_2 e^{-iJt/2}, b_2 = \alpha_1 \beta_1 e^{-iJt/2}, b_3 = (\Delta_+ e^{-i(\delta - J)t/2} + \Delta_- e^{i(\delta + J)t/2})/2$ and $b_4 = (\Delta_+ e^{-i(\delta - J)t/2} - \Delta_- e^{i(\delta + J)t/2})/2$. The eigenvalue equation for $\rho_1$ is

$$\tau^2 - (\rho_{11} + \rho_{00})\tau + (\rho_{11}\rho_{00} - |\rho_{10}|^2) = 0. \tag{12}$$

To achieve a swap operation, we must have a product state of spin 1 and 2 evolve into a product state, and the Schmidt number of the two-spin state cannot exceed one, which means that only one eigenvalue of the reduced density $\rho_1$ is nonvanishing, so $\rho_{11}\rho_{00} - |\rho_{10}|^2 = 0$. From Eq. (11) we have:

$$\begin{aligned} &\rho_{1,00}\rho_{1,11} - |\rho_{1,10}|^2 = |b_1 b_2 - b_3 b_4|^2 \\ &= \left| \begin{matrix} \frac{1}{2} e^{iJt} \{\sin[\delta t]\sin[\theta](\alpha_1^2 \beta_2^2 - \alpha_2^2 \beta_1^2) + 2\alpha_1 \alpha_2 \beta_1 \beta_2 (\cos[2Jt] - \cos[\delta t])\} \\ + i\{\sin[\delta t]\cos[\theta](\alpha_2^2 \beta_1^2 + \alpha_1^2 \beta_2^2) - 2\alpha_1 \alpha_2 \beta_1 \beta_2 \sin[2Jt]\} \end{matrix} \right|^2. \end{aligned} \tag{13}$$

Now consider the value of $t$ that makes Eq. (13) vanish. If $t$ depends on the initial state parameters $\alpha_1, \alpha_2$, $\beta_1$ and $\beta_2$, then for an unknown initial state the swap operation can't be realized. Hence $t$ must be independent of the initial state parameters. Therefore we must have $\delta t = k\pi$ and $2Jt = m\pi$, where $k, m = 0, \pm 1, \pm 2, \cdots$. Because $\delta = 2J\sqrt{1 + D^2}$, there is a $t$ that makes Eq. (13) vanish only when $\sqrt{1 + D^2}$ is a rational number. After a simple calculation, we know that only when $\sqrt{1 + D^2}$ is an odd rational number, the swap operation can be achieved, otherwise, the two-spin state return to the initial state with a phase shift. Table I gives the swap operation for $\sqrt{1 + D^2} = 1, 5, 9, \cdots$, and table II shows the results for $\sqrt{1 + D^2} = 3, 7, 11, \cdots$. We can see that the states of the two spins swap when $t=\frac{k\pi}{2J}$, ($k$ is an odd rational number) with an additional phase shift, so that the swap operation is achieved after the additional phase shift is corrected by a single-spin operation. It can also be found that the additional phase depends on the evolution time and the value of $\sqrt{1+D^2}$. Comparing Table II with Table I, we can see that the additional phase is appended with $e^{i\pi}$ for $\sqrt{1+D^2}=3,7,11,\cdots$. When $D=0$ and $\theta=0$, which corresponds to $\sqrt{1+D^2}=1$ in Table I, the model is reduced to a Heisenberg XXX one. From Table I, the swap operation is achieved with an additional phase shift $e^{-i\pi/8}$. This is in accord with the results in Refs. [11, 12].

## V. CONCLUSION

To summarize, we have investigated the mutual information and swap operation in a two-qubit model in the presence of the DM anisotropic antisymmetric interaction. We find that the mutual information of such a quantum channel is sensitive to the initial conditions, and the entanglement quality of the input states cannot enhance the mutual information of the quantum channel. The DM interaction can lead to higher mutual information, especially for the ferromagnetic case. A swap operation can be realized for some special DM coupling with an additional phase shift. The additional phase shift depends on the evolution in time and the DM coupling. If we treat the DM coupling to be zero, our calculation accords with the results of Ref. [20], in which the authors studied the swap gate from a geometric perspective.

## ACKNOWLEDGEMENT

This work is supported by the National Science Foundation of China under Grants No. 10604053.

Figure captions:

Fig.1: (Color online) Mutual information *I* as a function of the input signal entangled states ($\omega$, $\Gamma$) for *J*=1 and *T*=*J*/(2ln3). (a) *D*=0; (b) *D*=5. *T* is plotted in units of the Boltzmann constant *k*, and we work in units where *D* and *J* are dimensionless.

Fig.2: (Color online) Mutual information *I* as a function of *J* and *T* for $\omega=\Gamma=\pi/4$. (a) *D*=0; (b) *D*=2. *T* is plotted in units of the Boltzmann constant *k*, and we work in units where *D* and *J* are dimensionless.

Fig.3: (Color online) Mutual information *I* as a function of *T* for *J*=1. The left panel: *D*=0; the right panel: *D*=2. The solid line labels maximally entangled input states when $\omega=\Gamma=(n+1/4)\pi$ or $(n+3/4)\pi$ ( $n=0,1,2,\cdots$ ). The dotted line labels non-entangled input states when $\omega = \Gamma = n\pi$ or $(n+1/2)\pi$ ( $n = 0,1,2,\cdots$ ). *T* is plotted in units of the Boltzmann constant *k*, and we work in units where *D* and *J* are dimensionless.

Fig.4: (Color online) Mutual information *I* as a function of *D* for *J*=1 and *T*=*J*/(2ln3). The solid line labels maximally entangled input states when $\omega = \Gamma = (n+1/4)\pi$ or $(n+3/4)\pi$ ( $n = 0,1,2,\cdots$ ). The dotted line labels non-entangled input states when $\omega = \Gamma = n\pi$ or $(n+1/2)\pi$ ( $n = 0,1,2,\cdots$ ). *T* is plotted in units of the Boltzmann constant *k*, and we work in units where *D* and *J* are dimensionless.

Table captions:

Table I: The swap operation for $\sqrt{1+D^2} = 1,5,9,\cdots; n = 0,1,2,\cdots$.

Table II: The swap operation for $\sqrt{1+D^2} = 3,7,11,\cdots; n = 0,1,2,\cdots$.

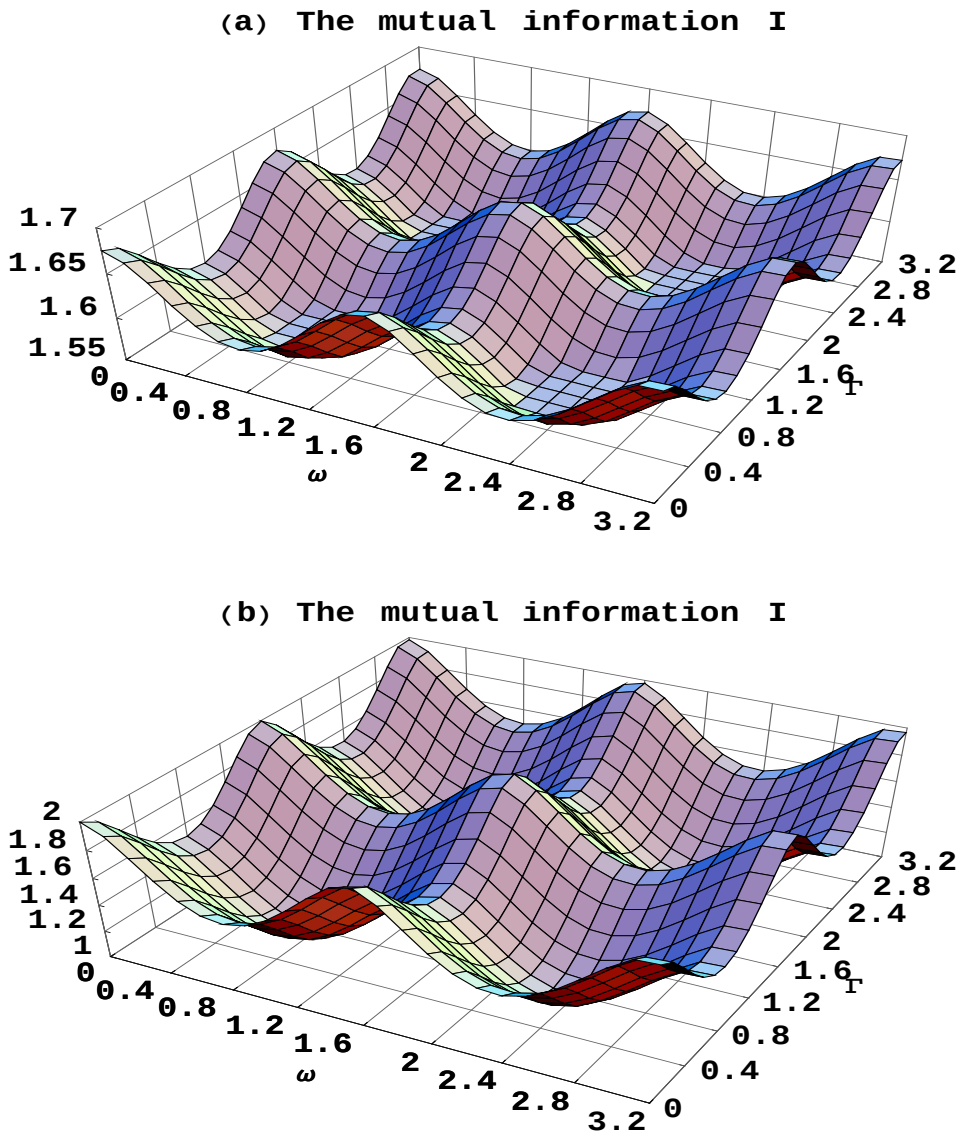

Fig.1

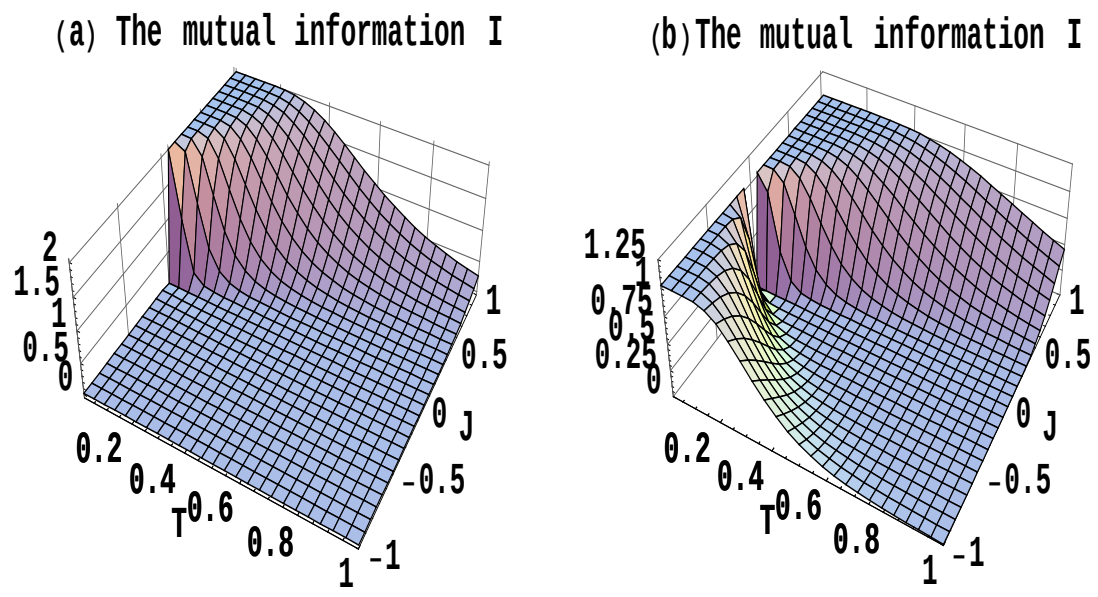

Fig.2

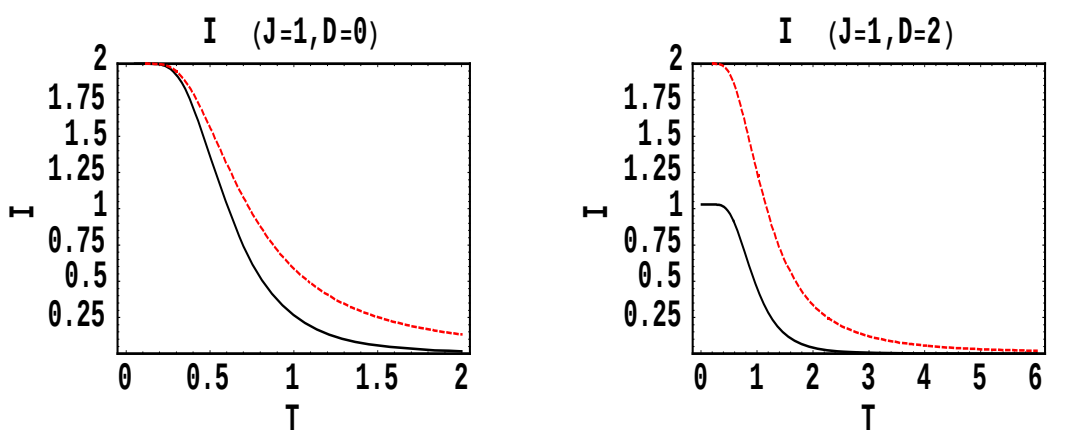

Fig.3

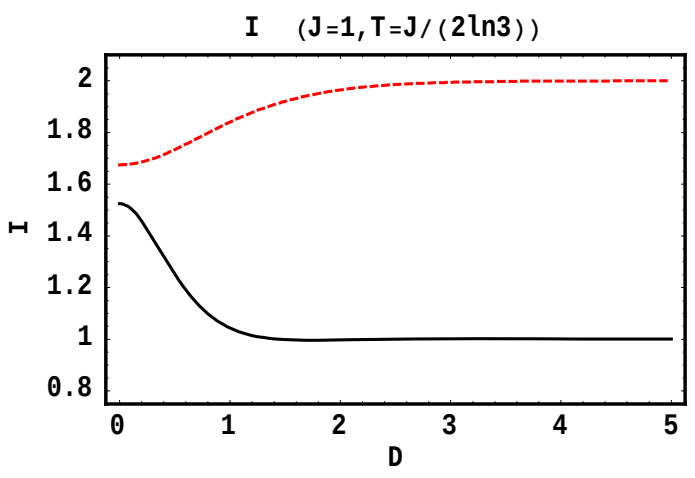

Fig.4

Table 1
The swap operation for $\sqrt{1+D^2} = 1, 5, 9, \ldots; n = 0, 1, 2, \ldots$

| $\psi(0) = (\alpha_1\|1\rangle + \alpha_2\|0\rangle) \otimes (\beta_1\|1\rangle + \beta_2\|0\rangle) = \binom{\alpha_1}{\alpha_2} \otimes \binom{\beta_1}{\beta_2}$ | | | |
|---|---|---|---|
| $t = \frac{\pi}{2J}$ | $t = \frac{3\pi}{2J}$ | $t = \frac{5\pi}{2J}$ | $t = \frac{7\pi}{2J}$ |
| $\psi(t) = \psi\left(t + \frac{4n\pi}{J}\right) = \begin{pmatrix} \beta_1 e^{-i\pi/8} \\ \beta_2 e^{-i[\pi/8+\theta]} \end{pmatrix} \otimes \begin{pmatrix} \alpha_1 e^{-i\pi/8} \\ \alpha_2 e^{-i[\pi/8-\theta]} \end{pmatrix}$ | $\psi(t) = \psi\left(t + \frac{4n\pi}{J}\right) = \begin{pmatrix} \beta_1 e^{-i3\pi/8} \\ \beta_2 e^{-i[3\pi/8+\theta]} \end{pmatrix} \otimes \begin{pmatrix} \alpha_1 e^{-i3\pi/8} \\ \alpha_2 e^{-i[3\pi/8-\theta]} \end{pmatrix}$ | $\psi(t) = \psi\left(t + \frac{4n\pi}{J}\right) = \begin{pmatrix} \beta_1 e^{i3\pi/8} \\ \beta_2 e^{i[3\pi/8-\theta]} \end{pmatrix} \otimes \begin{pmatrix} \alpha_1 e^{i3\pi/8} \\ \alpha_2 e^{i[3\pi/8+\theta]} \end{pmatrix}$ | $\psi(t) = \psi\left(t + \frac{4n\pi}{J}\right) = \begin{pmatrix} \beta_1 e^{i\pi/8} \\ \beta_2 e^{i[\pi/8-\theta]} \end{pmatrix} \otimes \begin{pmatrix} \alpha_1 e^{i\pi/8} \\ \alpha_2 e^{i[\pi/8+\theta]} \end{pmatrix}$ |

Table 2
The swap operation for $\sqrt{1+D^2} = 3, 7, 11, \ldots; n = 0, 1, 2, \ldots$

| $\psi(0) = (\alpha_1\|1\rangle + \alpha_2\|0\rangle) \otimes (\beta_1\|1\rangle + \beta_2\|0\rangle) = \binom{\alpha_1}{\alpha_2} \otimes \binom{\beta_1}{\beta_2}$ | | | |
|---|---|---|---|
| $t = \frac{\pi}{2J}$ | $t = \frac{3\pi}{2J}$ | $t = \frac{5\pi}{2J}$ | $t = \frac{7\pi}{2J}$ |
| $\psi(t) = \psi\left(t + \frac{4n\pi}{J}\right) = \begin{pmatrix} \beta_1 e^{-i\pi/8} \\ \beta_2 e^{-i\pi/8} e^{i[\pi+\theta]} \end{pmatrix} \otimes \begin{pmatrix} \alpha_1 e^{-i\pi/8} \\ \alpha_2 e^{-i\pi/8} e^{i[\pi+\theta]} \end{pmatrix}$ | $\psi(t) = \psi\left(t + \frac{4n\pi}{J}\right) = \begin{pmatrix} \beta_1 e^{-i3\pi/8} \\ \beta_2 e^{-i3\pi/8} e^{i[\pi-\theta]} \end{pmatrix} \otimes \begin{pmatrix} \alpha_1 e^{-i3\pi/8} \\ \alpha_2 e^{-i3\pi/8} e^{i[\pi+\theta]} \end{pmatrix}$ | $\psi(t) = \psi\left(t + \frac{4n\pi}{J}\right) = \begin{pmatrix} \beta_1 e^{i3\pi/8} \\ \beta_2 e^{i3\pi/8} e^{i[\pi-\theta]} \end{pmatrix} \otimes \begin{pmatrix} \alpha_1 e^{i3\pi/8} \\ \alpha_2 e^{i3\pi/8} e^{i[\pi+\theta]} \end{pmatrix}$ | $\psi(t) = \psi\left(t + \frac{4n\pi}{J}\right) = \begin{pmatrix} \beta_1 e^{i\pi/8} \\ \beta_2 e^{i\pi/8} e^{i[\pi-\theta]} \end{pmatrix} \otimes \begin{pmatrix} \alpha_1 e^{i\pi/8} \\ \alpha_2 e^{i\pi/8} e^{i[\pi+\theta]} \end{pmatrix}$ |